\documentclass[aip,jcp,amsmath,amssymb,reprint]{revtex4-1}
\usepackage{graphicx}
\usepackage{lipsum}
\usepackage{dcolumn}
\usepackage{amssymb}
\usepackage{bm}
\usepackage[utf8]{inputenc}
\usepackage[T1]{fontenc}
\usepackage{mathptmx}
\usepackage{etoolbox}
\makeatletter
\def\@email#1#2{%
 \endgroup
 \patchcmd{\titleblock@produce}
  {\frontmatter@RRAPformat}
  {\frontmatter@RRAPformat{\produce@RRAP{*#1\href{mailto:#2}{#2}}}\frontmatter@RRAPformat}
  {}{}
}%
\usepackage{physics}

\usepackage[normalem]{ulem}
\usepackage{xcolor}

\newcommand{\PHthree}{PH\textsubscript 3}

\makeatother
\begin{document}
%\preprint{AIP/123-QED}

\title{Fantastical Excited State Optimized Structures and Where to Find Them} % [MAYBE: Are your optimized excited state structures ``real'' and, if not, what are they?]}

 \author{Justin J. Talbot}
 \affiliation{Department of Chemistry, University of California Berkeley, California 94720, USA}
 \email{justin.talbot@berkeley.edu}

 \author{Juan E. Arias-Martinez}%
 \affiliation{Department of Chemistry, University of California Berkeley, California 94720, USA}
 \affiliation{Chemical Sciences Division, Lawrence Berkeley National Laboratory, Berkeley, California 94720, USA}

\author{Stephen J. Cotton}
\affiliation{Department of Chemistry, University of California Berkeley, California 94720, USA}
\email{stephenjcotton47@gmail.com}

\author{Martin Head-Gordon}
\affiliation{Department of Chemistry, University of California Berkeley, California 94720, USA}
\affiliation{Chemical Sciences Division, Lawrence Berkeley National Laboratory, Berkeley, California 94720, USA}
\email{mhg@cchem.berkeley.edu}

\date{\today}

\begin{abstract}
The quantum chemistry community has developed analytic forces for approximate electronic excited states to enable walking on excited state potential energy surfaces (PES). One can thereby computationally characterize excited state minima and saddle points. Always implicit in using this machinery is the fact that an excited state PES only exists within the realm of the Born-Oppenheimer approximation, where the nuclear and electronic degrees of freedom separate. This work demonstrates through \textit{ab initio} calculations and simple nonadiabatic dynamics that some excited state minimum structures are fantastical: they appear to exist as stable configurations only as a consequence of the PES construct, rather than being physically observable. One such case is the S$_2$ excited state of phosphine and a second case are local minima of a number of states of tris(bipyridine)ruthenium(II). Each fantastical structure exhibits an unphysically high predicted harmonic frequency and associated force constant. This fact can serve as a valuable diagnostic of when an optimized excited state structure is non-observable. Their origin lies in the coupling between different electronic states, and the resulting avoided crossings. The upper state may exhibit a minimum very close to the crossing, where the force constant relates to the strength of the electronic coupling rather than to any characteristic excited state vibration. Nonadiabatic dynamics results using a Landau-Zener model illustrate that fantastical excited state structures have extremely short lifetimes on the order of a few femtoseconds. Their appearance in a calculation signals the presence of a nearby avoided crossing or conical intersection through which the system will rapidly cross to a lower surface.
\end{abstract}

\maketitle

Quantum chemistry methods for electronic excited states have undergone tremendous development, with a toolbox ranging simple tractable methods such as single excitation configuration interaction (CIS) and time-dependent density functional theory (TDDFT) to increasingly accurate and sophisticated equation-of-motion coupled cluster (EOM-CC) methods, as well as designer excited state wavefunctions of the multiconfiguration or complete active space (CAS) type. All such methods define the $S^\text{th}$ excited state potential energy surface (PES), $E_S(\mathbf{R})$, in terms of the Cartesian nuclear coordinates, $\mathbf{R}$, as a consequence of the clamped nucleus (i.e. Born-Oppenheimer) approximation. The resulting force, $\mathbf{F}_S=-\frac{\partial E_S}{\partial \mathbf{R}}$, allows one to walk on the PES and optimize excited state structures or transition structures.\cite{pulay2014analytical,mustroph2016potential,charaf2013choosing,baiardi2013general,bousquet2013excited,wang2020accurate,page2003ground,send2011assessing,wiberg2005comparison} Great effort has therefore gone into evaluating forces associated with excited state energies from CIS,\cite{maurice1998single} TDDFT,\cite{liu2010parallel,furche2002adiabatic,nguyen2010analytical,chiba2006excited,petrenko2011efficient} EOM-CC,\cite{stanton1993many,stanton1994analytic,feng2019implementation} and CAS\cite{szalay2012multiconfiguration} The same surface-walking optimization algorithms used for ground states can then be applied.\cite{schlegel2011geometry,baker1986algorithm,baker1993techniques,bakken2002efficient} Such methods are available in standard electronic structure codes, such as the Q-Chem software\cite{epifanovsky2021software}, which is used for all calculations reported here.

In comparison to the ground state, walking along an excited state PES to locate optimized structures comes with additional challenges.\cite{closser2014simulations,serrano2005quantum,steinmetzer2021pysisyphus} Perhaps the primary difficulty is that the excited state PES is shaped by the coupling between nearby electronic states. This coupling can act to distort the PES forming both local minima and sharply-curved avoided crossings that are near each other. When a surface walking algorithm finds one of these local minima, the increased curvature of the PES from a nearby avoided crossing can cause one of the resulting vibrational frequencies to be unphysically and enormously high. However, as illustrated below, these local minima are not dynamically stable and thus the anomalous frequencies are not physically observable. Rather, they result from a breakdown in the Born-Oppenheimer approximation and are fantastical artifacts of the PES being a poor mathematical construct.

In the adiabatic representation, avoided crossings are relatively common on excited state potential energy surfaces; and the dynamics through avoided crossings are of interest in areas such as reaction and photochemistry,\cite{schuurman2018dynamics, levine2007isomerization, bernardi1996potential, ibele2020molecular, atkins2017trajectory} material science and design,\cite{valiev2018first, hasselbrink2006non, xie2022generalized, cheshire2020ultrafast} and in the development of new technologies.\cite{kaestner2015non, cavaliere2009nonadiabatic, long2017nonadiabatic, levine2019locality} Experimental techniques, such as time-resolved spectroscopies, are beginning to attain the sensitivity required to discern signatures of passage through avoided crossings.\cite{schuurman2022time, kjaer2019finding, schnappinger2022time, stolow2003time, timmers2019disentangling} Of equal importance, theoretical modeling serves a key role in our understanding of their impact in the aforementioned processes and areas of study. Despite progress, identifying these regions of the PES remains a challenge for quantum chemistry algorithms.\cite{domcke2004conical, fernandez2012identification, yarkony2012nonadiabatic,williams2023geometric} The brute force approach relies on the calculation of first-order derivative coupling vectors which, while effective, can amount to the dominant computational expense and render the procedure intractable for larger system sizes.\cite{talbot2022symmetric, richter2011sharc, plasser2016efficient, ryabinkin2015fast, pittner2009optimization} Alternative methods, such as carrying out quasi-classical and quantum molecular dynamics simulations\cite{jasper2006non, nelson2014nonadiabatic, talbot2022dynamic} and measuring the density of avoided crossings,\cite{obzhirov2023density} have been presented.  Methods to find the minimum energy crossing point (MECP) between two different states have also been developed,\cite{harvey1998singlet,epifanovsky2007direct} as the MECP plays an important role in non-radiative relaxation.

The purpose of this communication is to illustrate and explain how a standard analytical gradient-based geometry optimization procedure may identify molecular configurations as excited state minima which are not dynamically stable. A hallmark of such an instability is an anomalously large vibrational frequency and force constant. As some representative examples, we will present the excited state PESs of gas-phase phosphine ({\PHthree}) where a geometry optimization using EOM-CCSD theory converges to a molecular configuration that is near an avoided crossing. A harmonic vibrational analysis at this local minimum identifies a single normal mode with an unexpectedly strong force constant. Similarly and confirming this analysis, the vibrational dependence of the first-order derivative coupling illustrates that this normal mode connects the minimum to a nearby avoided crossing. Next, a TDDFT analysis of tris(bipyridine)ruthenium(II) (RuBPY) is presented where we see the same type of behavior occurring with multiple minima on multiply-coupled excited state PESs. This second example more clearly illustrates the relationship between the strength of the first-order derivative coupling and the second derivative of the excited state energy. Finally, a simple  dynamical analysis based on the ubiquitous Landau-Zener model of an electronically non-diabatic curve-crossing illustrates that these fantastical minima are, in fact, unstable and have exceptionally short lifetimes.

%PH3
\begin{figure}[t]
\centering
  \includegraphics[height=7cm]{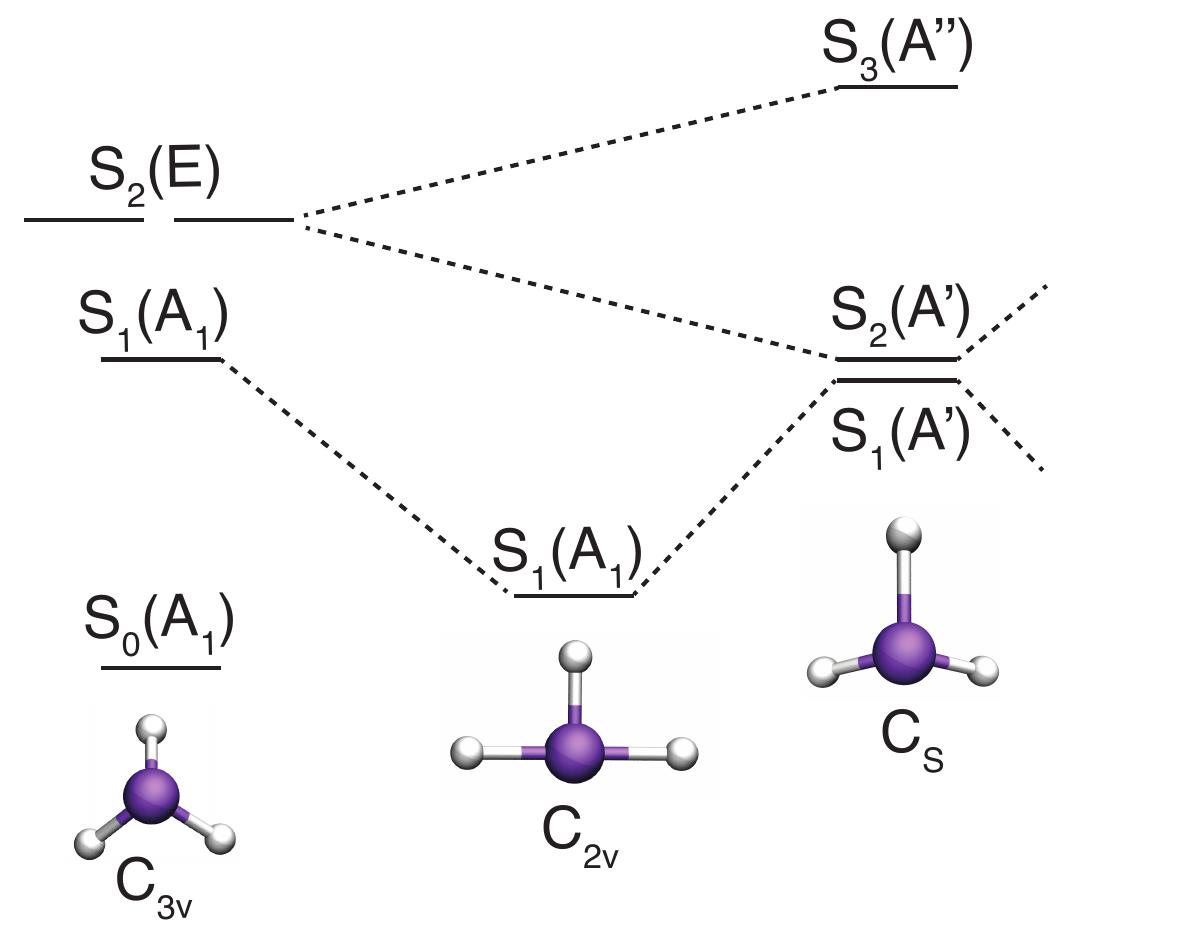}
  \caption{A schematic of the possible stationary points on the excited state PES's of {\PHthree}. At the C\textsubscript{3V} geometry, the S\textsubscript{1} excited state is A\textsubscript{1} symmetry and the S\textsubscript{2} excited state is a degenerate pair of E symmetry states. A geometry optimization splits the E electronic states into a lower energy S\textsubscript{2}(A') state and a higher energy S\textsubscript{3}(A'') state in the C\textsubscript{S} point group. The avoided crossing between the S\textsubscript{1}(A') and S\textsubscript{2}(A') surfaces occurs near the S\textsubscript{2}(A') minimum.}
  \label{Fig1}
\end{figure}

Our first example of a fantastical excited state minimum energy structure is {\PHthree} and because it is easily tractable with EOM-CCSD we are assured our analysis has not been skewed by a more modest level of electronic structure theory. A schematic of the energy ordering and minimum energy configurations is shown in Fig.~\ref{Fig1}. In its ground electronic state, {\PHthree} has A\textsubscript{1} symmetry in the C\textsubscript{3V} point group. {\PHthree's} first excited state has A\textsubscript{1} symmetry and its second excited state is a degenerate pair of states with E symmetry. An excitation into the A\textsubscript{1} state optimizes to a symmetric T-shaped configuration, also of A\textsubscript{1} symmetry, but in the C\textsubscript{2V} point group (see middle panel of Fig.~\ref{Fig1}). Upon excitation into one of the degenerate E states, {\PHthree} will lower symmetry to the C\textsubscript{S} point group by lengthening one of its P-H bonds and optimizing to a higher energy A'' minimum and a lower energy A' minimum energy configuration. An avoided crossing is formed between the S\textsubscript{1} and S\textsubscript{2} states which is near the A' minimum along a combined vibrational degree of freedom that both bends and breaks the planar symmetry.

\begin{figure*}
\centering
  \includegraphics[height=12cm]{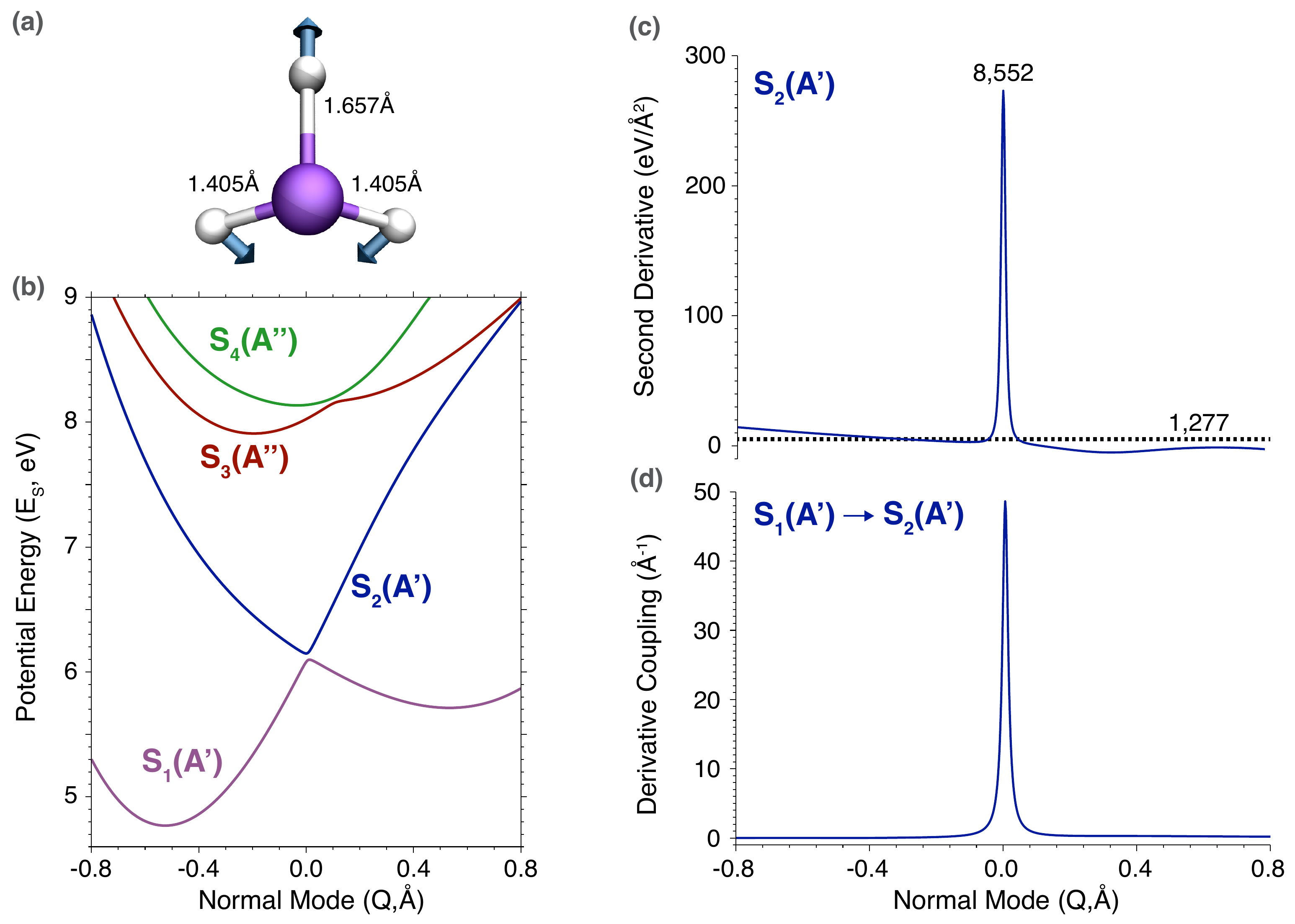}
  \caption{(\textbf{a}) The normal mode degree of freedom ($Q$) of {\PHthree} that connects the S\textsubscript{2}(A') minimum ($Q=0$) with the nearby avoided crossing. (\textbf{b}) The PES's as a function of this normal mode coordinate. (\textbf{c}) The second derivative of the S\textsubscript{2}(A') PES as a function of normal mode coordinate. The corresponding maximum vibrational frequency (cm$^{-1}$) is shown above the peak with the average frequency shown as a dashed line. (\textbf{d}) The first-order derivative coupling between the S\textsubscript{1}(A')-S\textsubscript{2}(A') states as a function of normal mode coordinate.   
  }
  \label{Fig2}
\end{figure*}

Fig.~\ref{Fig2} presents an EOM-EE-CCSD/aug-cc-pVDZ analysis of the PESs of four states of {\PHthree} as a function of the normal mode coordinate ($Q$). Also shown, as a function of $Q$, is the second derivative of the excited S\textsubscript{2}(A') state energy [panel~(c)] (giving the fantastical force constant) along with the first-order derivative coupling between this state and the lower-in-energy S\textsubscript 1(A') state [panel~(d)]. The anomalous normal mode [panel~(a)] consists of a bending motion of the equatorial hydrogen atoms that have both in- and out-of-plane character with a stretching of the axial P-H bond. The low-lying excited state PESs [panel~(b)] consist of two singlet A' states followed by two higher-energy  A'' states. A harmonic vibrational analysis at the fantastical S\textsubscript{2}(A') minimum predicts that the vibrational frequency is $6,884$~cm$^{-1}$ which is $\approx~5$ times greater than the average vibration frequency of $1,277$~cm$^{-1}$. Remarkably, at the point where the first-order derivative coupling reaches a maximum, which occurs just off the minimum at $Q\approx0.006$~{\AA}, the second derivative of the S\textsubscript{2}(A') state energy reaches a maximum. The vibrational frequency at this point on the PES is even greater ($8,552$~cm$^{-1}$) when compared to the frequency at the minimum. Walking along the PES of this normal mode connects the minimum energy structure to a nearby avoided crossing as evidenced by both the increased second derivative along this degree of freedom and the increased first-order derivative coupling. The curvature of the S\textsubscript{2}(A') PES along this direction is significantly changing as a result of the nearby avoided crossing. 

%RuBpy
\begin{figure*}[t]
\centering
  \includegraphics[height=11cm]{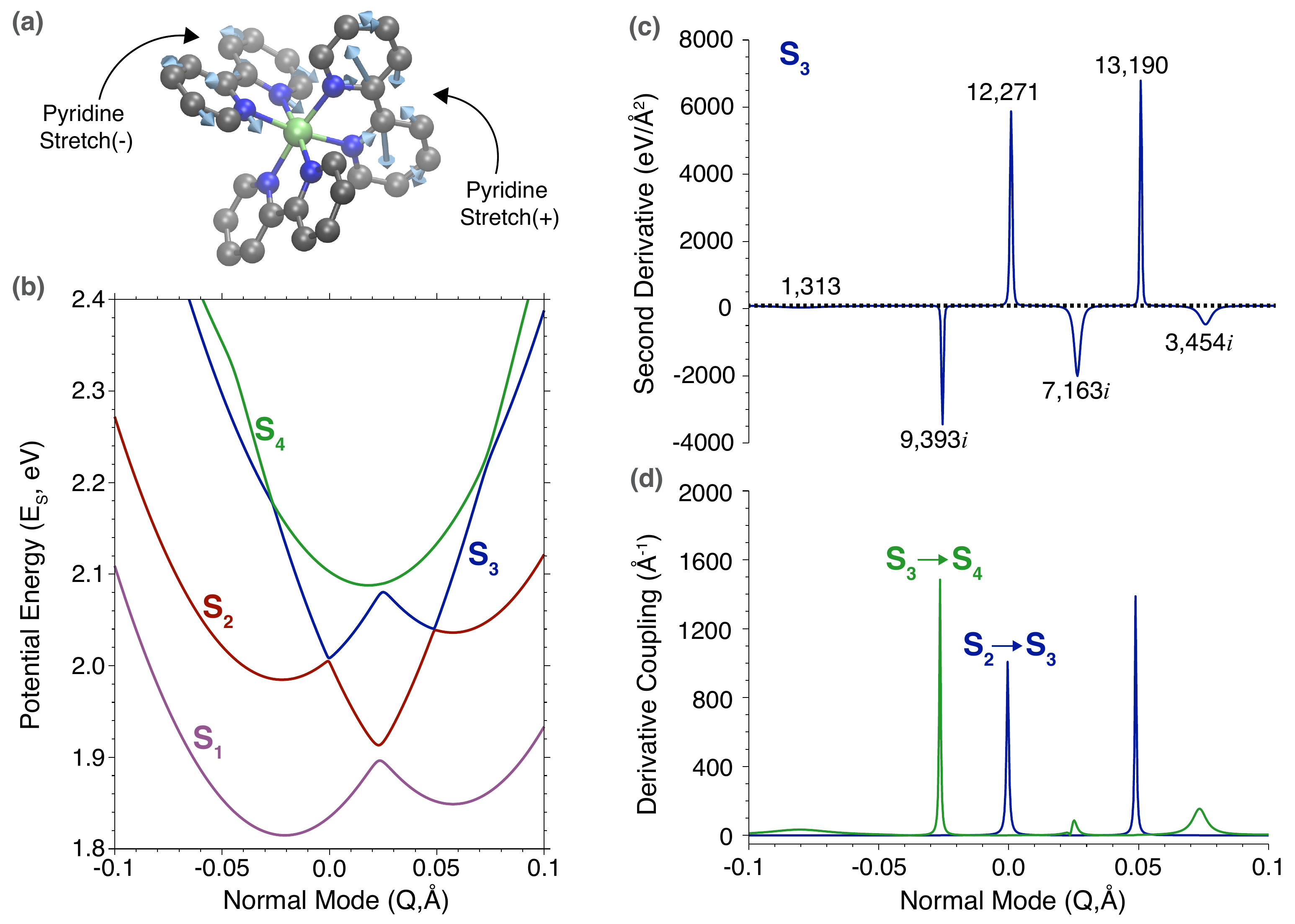}
  \caption{(\textbf{a}) The normal mode degree of freedom ($Q$) of RuBPY that connects the S\textsubscript{3} minimum ($Q=0$) with the nearest avoided crossings. The normal mode consists of a symmetric pyridine-pyridine streching motion on two of the bipyridine ligands oscillating in- and out-of-phase.(\textbf{b}) The PES's of the lowest four excited states as a function of $Q$. (\textbf{c}) The second derivative of the S\textsubscript{3} excited state energy as a function of $Q$ with the local maximum and minimum vibrational frequencies (cm$^{-1}$) displayed above the peaks. The average vibrational frequency is shown as a dashed line. (\textbf{d}) The first-order derivative coupling between the S\textsubscript{3}-S\textsubscript{4} (green) and S\textsubscript{2}-S\textsubscript{3} (blue) states as a function of $Q$.  
  }
  \label{Fig3}
\end{figure*}

Our second example, shown in Fig.~\ref{Fig3}, treats the B3LYP/def2-SVP excited state PESs of RuBPY. The analysis of this fantastical excited state landscape consists of four PESs [panel~(b)] connected by normal mode $Q$ with five avoided crossings occurring between them. A schematic of the anomalous normal mode degree of freedom [panel~(a)] involves symmetric pyridine-pyridine stretches on two of the bipyriding ligands. Walking along a relatively small displacement ($\pm 0.1${\AA}) of this vibrational degree of freedom on the S\textsubscript{3} excited state reveals that when the first-order derivative coupling [panel~(d)] reaches a maximum, the second derivative of the S\textsubscript{3} energy [panel~(c)] reaches an inflection point. A harmonic vibrational analysis at the fantastical S\textsubscript{3} minimum predicts that the vibrational frequency is $11,801$~cm$^{-1}$ which is $\approx 9$ times higher than the average vibrational frequency of $1,313$~cm$^{-1}$. At the nearest avoided crossing ($Q\approx0.0004$~{\AA}), the predicted vibrational frequency is an enormous $12,271$~cm$^{-1}$. Similar to the excited states of {\PHthree}, what is found by geometry optimizing the S\textsubscript{3} excited state of RuBPY are potential minima with anomalously large vibrational frequencies and force constants coincident with the presence of a nearby avoided crossings with strong first-order derivative couplings. The presence of multiple avoided crossing illustrates that strong fluctuations in the curvature of the PES are not unique to avoided crossings near minimum energy configurations but are indicative of the avoided crossings themselves. The fantastical predicted vibrational frequencies at the minimum are simply an artifact of the nearby avoided crossing.

%L-Z probability
These examples are more fully understood through a simple dynamical analysis based on the venerable Landau-Zener model,\cite{zener1932non,rubbmark1981dynamical,tully2012perspective} which illustrates that an excited state PES minimum having such an anomalously large force constant will have an extremely short lifetime and thus cannot truly be considered a stable point on an excited state PES. Fig.~\ref{Fig2} and Fig.~\ref{Fig3} shows for {\PHthree} and RuBPY that, in the adiabatic representation, the narrowly avoided crossings and high vibrational frequencies near these minimum energy geometries are accompanied by strong first-order derivative couplings. In the alternative view of a strictly diabatic representation however, these regions large first-derivative couplings correspond to weak coupling between the diabatic states. The Landau-Zener model presents the simplified case of two linear diabatic potential energy surfaces as a function of a single vibrational degree of freedom, which cross at $Q=0$, and have a constant coupling matrix element between them. After transforming from the diabatic basis of the Landau-Zener model into the adiabatic basis, the force constant at the minimum ($Q=0$) of the upper adiabatic state $E_+$ is:

\begin{equation}
  \frac{\partial^2E_{+}}{\partial Q^2}\Bigr|_{Q=0} = \frac{(G_I-G_J)^2}{4\Delta},
  \label{eqFC}
\end{equation} 
\\
\noindent where $G_I$ and $G_J$ denote the slopes of the diabatic potential energy surfaces and $\Delta$ is the constant diabatic coupling matrix element. As shown in Eq.~\ref{eqFC}, when the diabatic coupling is weak (e.g. near zero), the force constant of the upper adiabatic state is near infinite. If one adopts the Landau-Zener model's assumption of a constant nuclear velocity $\dot{Q}$ through this avoided crossing then the standard Landau-Zener population of the upper adiabatic state is: 

\begin{equation}
  P = 1 - \exp\bigg(\frac{-2\pi\Delta^2}{\hbar \dot{Q} |G_I - G_J|}\bigg),
  \label{eqLZ}
\end{equation} 
\\
\noindent where P denotes the population loss of $E_+$ after a \emph{single} pass through the avoided crossing region.

%Landau-Zener analysis
\begin{figure}[t]
\centering
  \includegraphics[height=11cm]{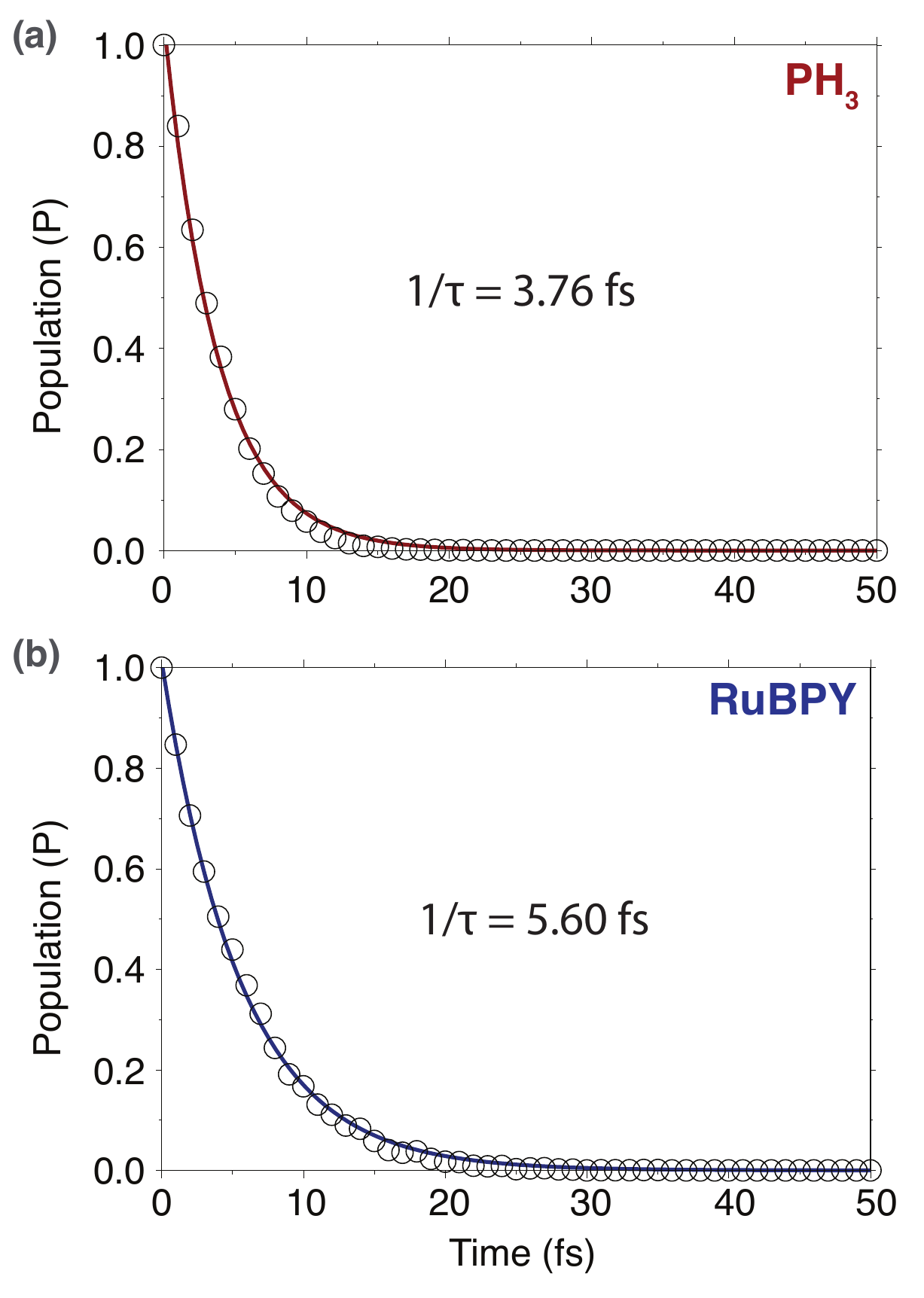}
  \caption{The average Landau-Zener populations (after considering re-crossings) of the upper adiabatic state as a function of time. Approximately 10,000 trajectories were sampled from the linearized PES's ground vibrational state distributions. (a) The population decay of the S\textsubscript{2}(A') excited state of {\PHthree} as a function of time. (b) The population decay of the S\textsubscript{3} excited state of RuBPY as a function of time. The lifetime $1/\tau$ is shown in the inset.  
  }
  \label{Fig4}
\end{figure}

The Landau-Zener model was used to predict the lifetimes of the fantastical minimum energy configurations of {\PHthree} and RuBPY. For this analysis, the PES's were fit to linear functions and the diabatic coupling $\Delta$ was calculated from the energy splittings of the adiabatic states at each minimum. 
In order to include the effects of re-crossing, classical molecular dynamics trajectories on the linear PES's were employed with the crossing times, and the resulting velocities at the crossing point, solved for analytically. Approximately 10,000 trajectories were initialized with positions and momenta sampled from the vibrational ground state probability density of the corresponding linearized PES's. The velocity $\dot{Q}$ inserted into Eq.~\ref{eqLZ} and the results were averaged. Further details of the linear fit and the molecular dynamics approach can be found in the supporting information. 

The population decay as a function of time is shown in Fig.~\ref{Fig4}. Fitting these curves to a first-order exponential function reveals that these fantastical minimum energy structures are incredibly short-lived. The average lifetime of the S\textsubscript{2}(A') minimum energy configuration of {\PHthree} is only $3.76$ fs. Likewise, the S\textsubscript{3} minimum energy configuration of RuBPY is only $5.60$ fs. It's important to note that these are conservative estimates of the lifetimes since the dynamics were initialized from the vibrational ground state. Realistically, as the avoided crossings are approached from further away (e.g. the Franck-Condon region), the nuclear velocities would likely be much greater and the population would decay even faster.    
 
%Conclusions
In summary, this work serves to highlight striking features of excited state potential energy surfaces which are associated with walking downhill to a minimum on an upper state, $j$, that is near an avoided crossing or seam of intersection with a lower state, $(j-1)$. We know abstractly that the very concept of a potential energy surface, $E_j(\mathbf{R})$ is an idealization of the strict Born-Oppenheimer separation of electron motion ($\mathbf{r}$) from nuclear motion ($\mathbf{R}$), which yields the electronic Schr\"odinger equation, $\hat{H}_e \Psi_j(\mathbf{r};\mathbf{R}) = E_j(\mathbf{R}) \Psi_j(\mathbf{r};\mathbf{R})$. For most purposes, the PES is an incredibly valuable as well as quantitatively valid construct, because it gives rise to fundamental chemical concepts such as molecular structure, and thus reaction mechanism as the passage between a sequence of such structures. However, when we walk on $E_j(\mathbf{R})$ and obtain optimized structures and their properties, such as vibrational frequencies, we tend to assume those results are valid while forgetting that the PES itself is an idealization. Our discovery of non-observable (``fantastical'') optimized excited state structures that are associated with the upper surface of avoided crossings highlights the need to not unconditionally accept well-characterized excited state PES minima as physically meaningful. 

In the case of the fantastical excited state minimum structures we reported, there is a clear signature of Born-Oppenheimer breakdown in the form of unphysical artifacts in the harmonic vibrational frequency analysis. Specifically there is an enormously high frequency (and force constant) that far exceeds normal values. This ``mega-mode'' arises from the very sharp form of the avoided crossing on the upper state, as is visually evident, for example in the extreme curvature associated with the lowest S$_2$ and S$_3$ structures in Figure \ref{Fig3}(b). Furthermore, although it depends on the strength of the electronic coupling matrix elements, these fantastical minimum energy configurations are typically unstable with incredibly short lifetimes. To demonstrate the exceptionally short-lived occurrences of these metastable configurations a Landau-Zener analysis was presented. For the representative examples of {\PHthree} and RuBPY, a conservative analysis predicted lifetimes of less than $10$ fs out of the excited states of interest. Clearly, these minimum energy configurations are unphysical artifacts of the Born-Oppenheimer approximation.  

We therefore recommend examining the results of excited state PES optimizations to make sure the resulting structures are not fantastical. If a fantastical structure is discovered, with its associated mega-mode, it serves as an indicator of a very near-by seam of crossing, rather than a conventional minimum structure.

\section{Acknowledgments}

\noindent This work is supported by the Director, Office of Science, Office of Basic Energy Sciences of the US Department of Energy under contract No. DE-AC02-05CH11231. This work is supported by the National Science Foundation under grant number CHE-1856707. This research used resources of the National Energy Research Scientific Computing Center (NERSC), a U.S. Department of Energy Office of Science User Facility located at Lawrence Berkeley National Laboratory, operated under Contract No. DE-AC02-05CH11231 using NERSC award BES-ERCAP0025080.

\section{Conflict of Interest}

\noindent MHG is a part-owner of Q-Chem Inc, whose software was used for the developments and calculations reported here. 

\section{Author Contributions}

\noindent \textbf{Justin J. Talbot}: Conceptualization (equal); Analysis (lead); Methodology (lead); Writing - original draft (lead); Writing - review \& editing (equal). \textbf{Juan E. Arias-Martinez}: Conceptualization (equal); Analysis (supporting); Methodology (supporting); Writing - review \& editing (supporting). \textbf{Stephen J. Cotton}: Conceptualization (equal); Analysis (equal); Methodology (equal); Funding Acquisition (lead); Writing - review \& editing (supporting). \textbf{Martin Head-Gordon}: Conceptualization (equal); Analysis (supporting); Methodology (equal); Funding Acquisition (lead); Writing - review \& editing (equal).

\section{Data Availability}

\noindent The data that supports the findings in this study is available from the corresponding author upon request. Q-Chem frequency analysis files for the \PHthree and RuBPY examples are provided in the supporting information.

\section{References}
\bibliography{MMode}

\end{document}

\maketitle
\newpage

\section{The Landau-Zener Model}
The response of the second derivative of the excited energy to a nearby avoided crossing is best explained using a simple diabatic model. The model consists of two electronic states with potential energy surfaces (PES) that are linear functions of a single vibrational degree of freedom ($Q$) and which cross at $Q=0$. This two-state diabatic system is coupled by a constant coupling term $\Delta$. The Hamiltonian matrix elements of this model system are:
\begin{subequations}\label{e:Ham}
\begin{align}
    H_{11}(Q) &= G_{I}Q \\
    H_{22}(Q) &= G_{J}Q\\
    H_{12}(Q) & = \Delta,
\end{align}
\end{subequations}

\noindent where $G_{I}$ and $G_{J}$ naturally have opposite sign (for this analysis $G_I$ is negative and $G_J$ is positive) in order to enforce the crossing condition and denotes the slopes of the diabatic potential energy surfaces for states $I$ and $J$ respectively. The diabatic coupling $\Delta$ is constant. The eigenvalues (i.e. the adiabatic state energies) of this Hamiltonian are:
\\
\begin{equation}
    E_{\pm}(Q) = (G_{I} + G_{J})Q + \frac{1}{2}\sqrt{Q^2(G_{I}-G_{J})^2+4\Delta^2}
\end{equation}
\\
\noindent which are functions of $Q$ and depend on the sum and difference of the slopes of the diabatic surfaces and the coupling strength. The second derivative of the energy with respect to $Q$ for the upper adiabatic state $E_+$ is:
\\
\begin{equation}
  \frac{\partial^2E_{+}(Q)}{\partial Q^2} = \frac{2 \Delta^2(G_I - G_J)^2}{(Q^2(G_I-G_J)^2 + 4\Delta^2)^{3/2}},
\end{equation}
\\
\noindent which depends inversely on the coupling $\Delta$. Clearly, the upper adiabatic state $E_{+}(Q)$ reaches a minimum at $Q=0$ and the second derivative (i.e. the force constant) evaluated at this minimum is:
\\
\begin{equation}
  \frac{\partial^2E_{+}}{\partial Q^2}\Bigr|_{Q=0} = \frac{(G_I-G_J)^2}{4\Delta},
\end{equation} 
\\
\noindent which depends only on the difference between the slopes of the diabatic surfaces and inversely on $\Delta$. In the limit that the coupling goes to zero (i.e. $\Delta \rightarrow 0$), the force constant goes to infinity.

\begin{figure*}
\centering
  \includegraphics[height=10.5cm]{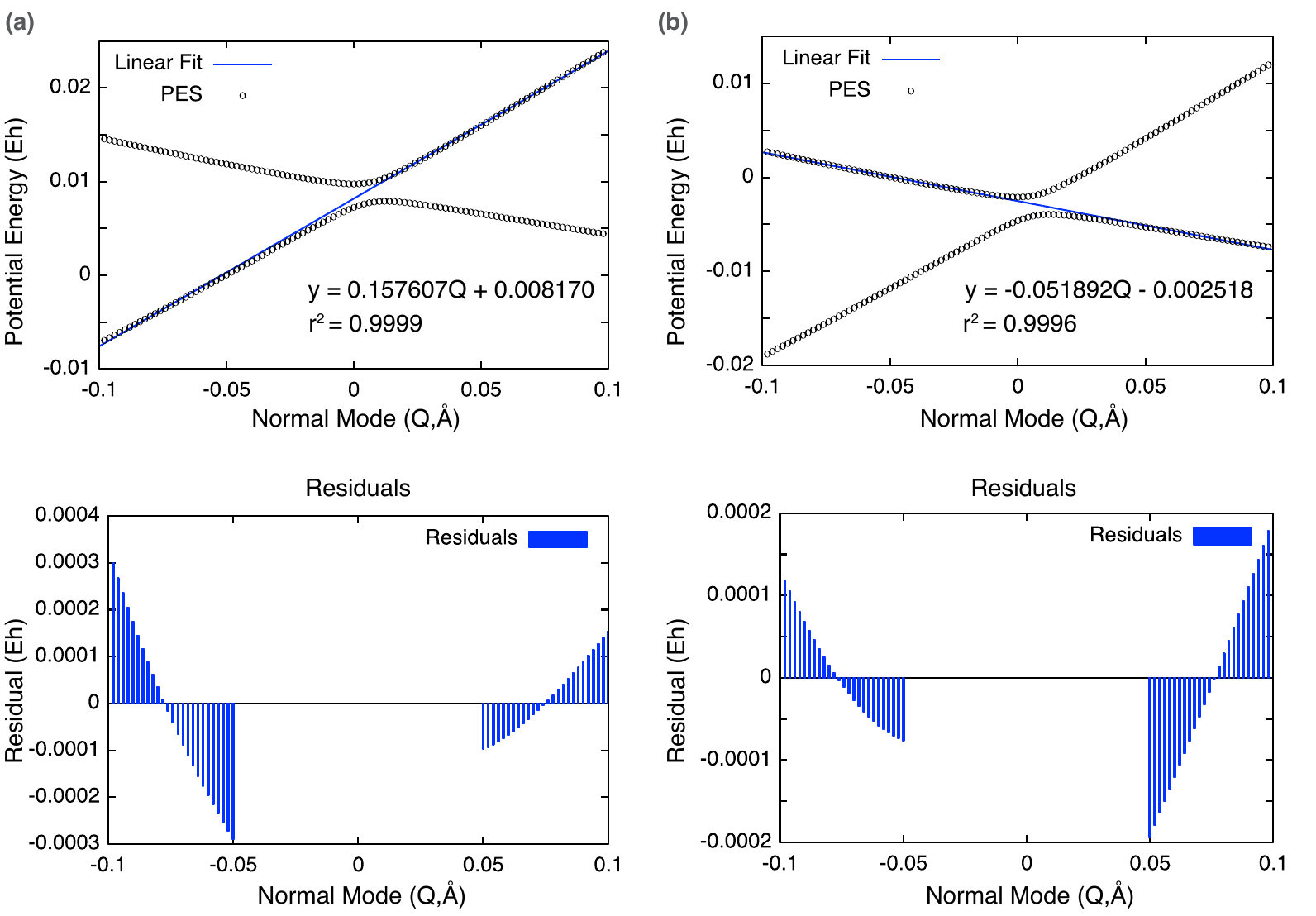}
  \caption{Linear fit analysis for PH3 (a) The positive linear fit (blue) as a function of normal mode displacement $Q$. The residual plot, in the fit region, is shown in the lower panel. (b) The negative linear fit (blue) as a function of normal mode displacement. The residual plot is shown in the lower panel.}
  \label{FigS1}
\end{figure*}

\begin{figure*}
\centering
  \includegraphics[height=10.5cm]{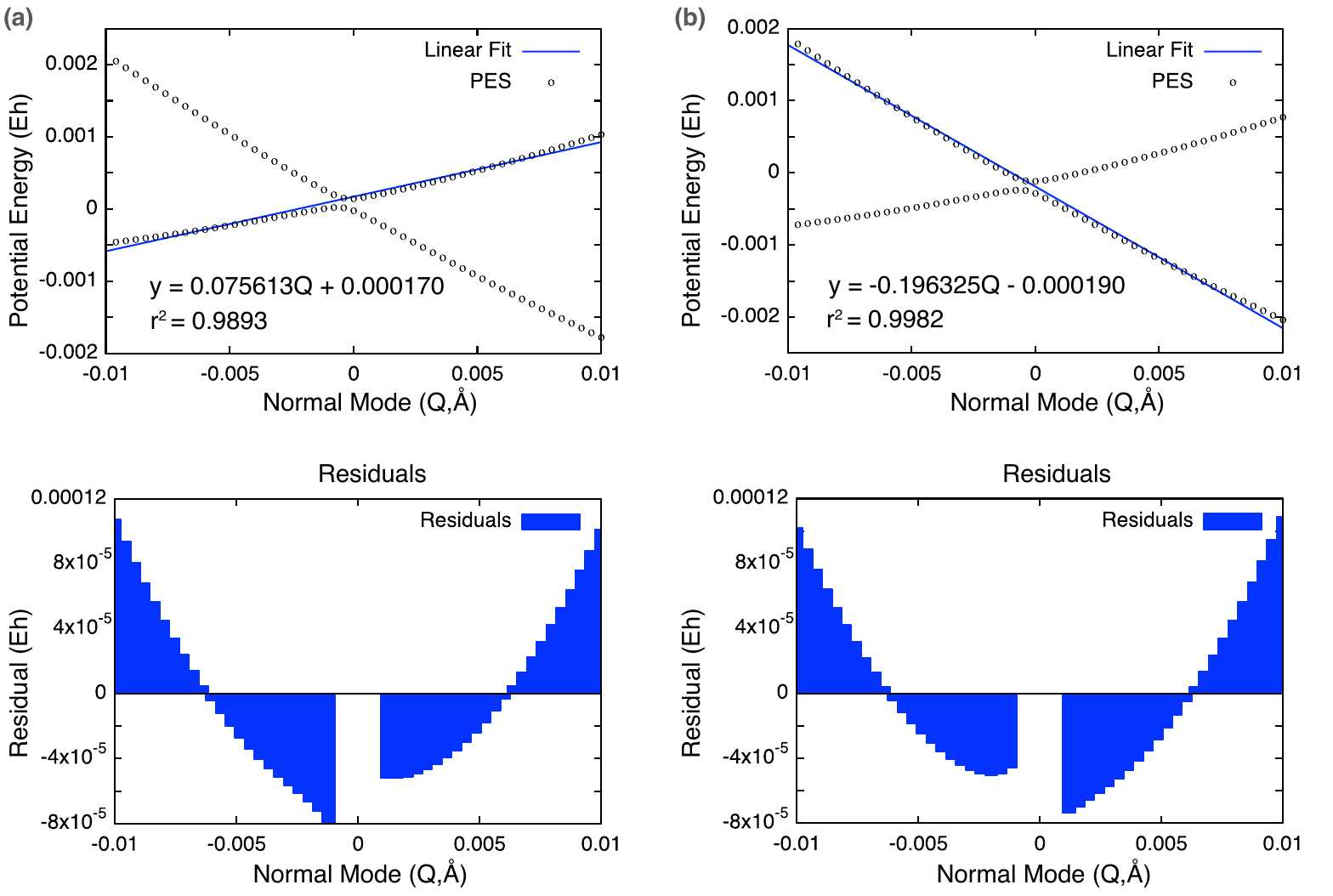}
   \caption{Linear fit analysis for RuBPY (a) The positive linear fit (blue) as a function of normal mode displacement $Q$. The residual plot, in the fit region, is shown in the lower panel. (b) The negative linear fit (blue) as a function of normal mode displacement. The residual plot is shown in the lower panel.}
  \label{FigS2}
\end{figure*}

The slopes, $G_I$ and $G_J$, were obtained by fitting the adiabatic potential energy surfaces around the fantastical minimum energy configurations in Fig.~2 and Fig.~3 of the main text. The linear fit analysis for PH\textsubscript{3} is shown in Fig.~\ref{FigS1} and the linear fit analysis for RuBPY is shown in Fig.~\ref{FigS2}. Some points around the minimum were excluded to get a better fit and these points are highlighted in the residual plots. For PH\textsubscript{3} all points between $\pm 0.05${\AA} were excluded and all points between $\pm 0.0005${\AA} were excluded for RuBPY. Goodness of fit was confirmed by the calculated $r^2$ values which were 0.9999 for PH\textsubscript{3} and 0.9996 for RuBPY. 

\begin{figure*}
\centering
  \includegraphics[height=10.5cm]{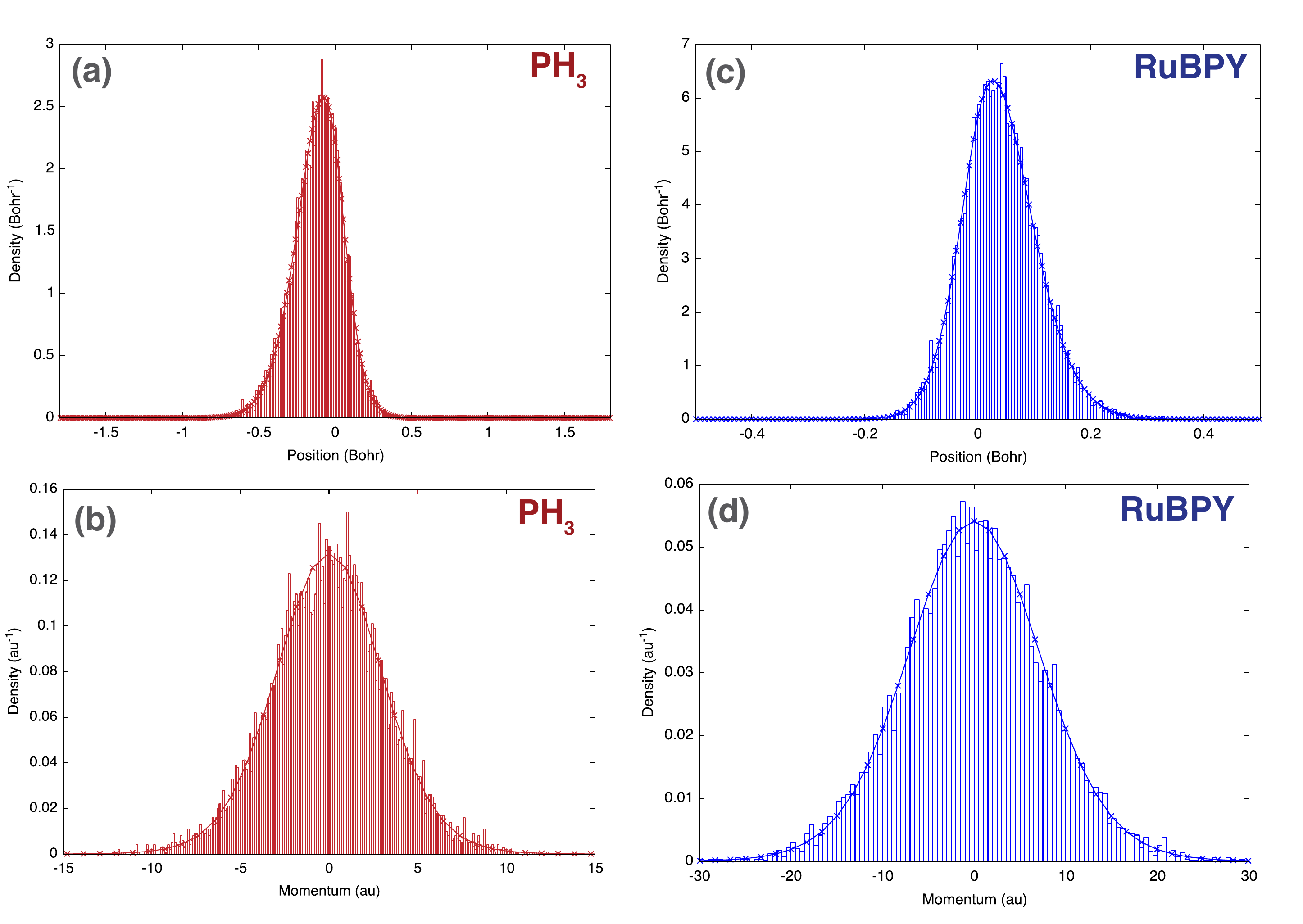}
  \caption{(a) The sampled initial positions (Bohr) for the PH\textsubscript{3} trajectories. (b) The sampled initial momenta (au) for the PH\textsubscript{3} trajectories. (c) The sampled initial positions (Bohr) for the RuBPY trajectories. (d) The sampled initial momenta (au) for the RuBPY trajectories. The eigensolver results are shown as a solid line and the samples are binned in the histograms. For each plot 10,000 positions/momenta were sampled.}
  \label{FigS3}
\end{figure*}

Using the slopes $G_I$ (right panel of Fig.~\ref{FigS2} and Fig.~\ref{FigS3}) and $G_J$ (left panel of Fig.~\ref{FigS2} and Fig.~\ref{FigS3}) molecular dynamics trajectories were employed on the upper diabatic state PES's. Since the slopes naturally have opposite sign, the upper state separates into a piecewise PES: 
\begin{subequations}\label{e:V}
\begin{align}
    E_L(Q) &= G_{I}Q \qquad Q<0 \\
    E_R(Q) &= G_{J}Q \qquad Q>0. 
\end{align}
\end{subequations}
\noindent Since the forces $\mathbf{F}_{L/R}=-\frac{\partial E_{L/R}}{\partial Q}$ are constant on a PES of this form, the equations of motion can be derived analytically:   
\begin{subequations}\label{e:EOM}
\begin{align}
    Q(t) &= Q_0 + V_0t + \frac{1}{2}at^2 \\
    \dot{Q}(t) &= V_0 + at,
\end{align}
\end{subequations}
\noindent where $Q_0$ is the initial position and $V_0$ is the initial velocity. Since the slopes to the left and right of $Q=0$ are different, the acceleration ($a$) also separates into a piecewise form:
\begin{subequations}\label{e:acc}
\begin{align}
    a_L &= -\frac{G_{I}}{\mu} \qquad Q<0 \\
    a_R &= -\frac{G_{J}}{\mu} \qquad Q>0,
\end{align}
\end{subequations}
\noindent where $\mu$ is the normal mode reduced mass. Calculating the population decay as a function of time requires the following properties that separate into two distinct cases which depend on the accelerations ($a_R$,$a_L$) initial position ($Q_0$), and initial velocity ($V_0$):
$$
Q_0\rightarrow Positive=
\begin{cases}
\dot{Q}_c = -\sqrt{V_0^2-2a_RQ_0}, \\
t_0 = \frac{-V_0+\dot{Q}_c}{a_R},\\
t_{2n+1} = t_{2n} - \frac{2\dot{Q}_c}{a_L}, \\
t_{2n} = t_{2n-1} + \frac{2\dot{Q}_c}{a_R},
\end{cases}
$$
and
$$
Q_0\rightarrow Negative=
\begin{cases}
\dot{Q}_c = \sqrt{V_0^2-2a_LQ_0}, \\
t_0 = \frac{-V_0+\dot{Q}_c}{a_L},\\
t_{2n+1} = t_{2n} - \frac{2\dot{Q}_c}{a_R}, \\
t_{2n} = t_{2n-1} + \frac{2\dot{Q}_c}{a_L},
\end{cases}
$$
\\
\noindent where the $n^{th}$ crossing time is denoted by $t_n$ (i.e. the $n^{th}$ time at which the system passes through the crossing point) and the corresponding velocity at the crossing point is denoted by $\dot{Q}_c$ (which is constant for a given set of initial conditions). With these values in hand, the population decays according to:
\begin{equation}
   P(t_n) = \bigg(P_{LZ}\bigg)^{n+1}
\end{equation}
\noindent where the Landau-Zener transition probability $P_{LZ}$ is defined as:
\begin{equation}
  P_{LZ} \equiv 1 - \exp\bigg(\frac{-2\pi\Delta^2}{\hbar |\dot{Q}_c(G_I - G_J)|}\bigg).
  \label{eqLZ}
\end{equation} 

\noindent For PH\textsubscript{3} and RuBPY, the reduced masses ($\mu$) are $1.0115$~amu and $10.6125$~amu respectively. In order to sample initial positions and momenta from the ground state vibrational probability density corresponding to the PES's in Eq.~{\ref{e:V}, the vibrational Schr\"{o}dinger equation was solved numerically using a Fourier grid variational eigensolver.\cite{marston1989fourier} The position and momentum distributions are shown in Fig.~\ref{FigS3}. The initial positions and momenta (10,000 for each system) were obtained by Monte-Carlo sampling of the these distributions. The lifetime estimates $1/\tau$ were obtained by fitting the following exponential function: 
\begin{equation}
   P(t) = A\exp(-t/\tau),
\end{equation}
to the averaged populations as a function of time ($t_n$, $n\rightarrow[0 \cdots 10]$). The averaging was performed over the initial $50$ fs in $1$ fs time intervals and the resulting populations were renormalized such that the population at time $t=0$ was set to unity (i.e. the averaged populations were divided by the average population at time $t=0$). 

\section{Additional Computational Details}

The second derivatives of the S\textsubscript{2}(A') excited state PES along the PH\textsubscript{3} normal mode degree of freedom (Fig.~2c in the main text) was calculated by finite difference of the analytic excited state gradient after projecting along the normal mode coordinate. The first-order derivative couplings, between the S\textsubscript{1}(A') and S\textsubscript{2}(A') states, were evaluated using Szalay's approach (as is the standard in Q-Chem) and similarly projected onto the normal mode coordinate.\cite{faraji2018calculations} The step size for both the finite difference second derivatives and the first-order derivative couplings was $0.002${\AA}.

The second derivative of the S\textsubscript{3} excited state energy along the RuBPY normal mode was also calculated using a finite difference approximation of the analytic gradient with a $0.0004$~{\AA} step size and projected onto the normal mode coordinate. The first-order derivative couplings were evaluated analytically using the methodology outlined in Ref.~\cite{talbot2022symmetric} and similarly projected onto the normal mode coordinate. The corresponding def2-ECP pseudopotential for the def2-SVP basis set was employed and represented 28 core electrons on the ruthenium atom. 

\printbibliography
%\bibliography{MMode}